\documentclass[traditabstract]{aa}

\usepackage{graphicx}
\usepackage{txfonts}
\usepackage{lipsum}
\usepackage{subcaption}         
\usepackage{lscape}             
\usepackage{placeins}           
\usepackage{chemformula}
\usepackage{natbib}

\usepackage[colorlinks=true, allcolors=blue]{hyperref}

\usepackage[utf8]{inputenc}
\usepackage{amssymb}
\usepackage{color}
\usepackage{amstext}
\usepackage{comment}
\usepackage[normalem]{ulem}
\usepackage{tikz}

\begin{document}

   \title{Molecular mapping of an exoplanet with JWST: Detection of \ch{NH3} and \ch{HCN} in the temperate super-Jupiter Epsilon Indi Ab}

 \titlerunning{Molecular mapping of an exoplanet with JWST}

   \author{R. Kravtchenko\inst{1,2,3}
        \and O. Berné\inst{1}
        \and I. Schroetter\inst{1}
        \and P. Amiot\inst{1}
        \and F. Debras\inst{1}
        }

    \institute{
        Institut de Recherche en Astrophysique et Planétologie, Université de Toulouse, 9 Av. du Colonel Roche, 31400, Toulouse, France
    \and
        École polytechnique, Institut Polytechnique de Paris, Route de Saclay, 91120 Palaiseau, France       
    \and
        Observatoire de Paris, 77 Av. Denfert Rochereau, 75014, Paris, France
        }
   \date{Received September 21, 2026}

\abstract
{Epsilon Indi Ab is currently the coldest and one of the closest directly imaged exoplanets ($T_{\rm eff} = 275$ K, d=3.6 pc). It offers a unique opportunity to test our understanding of the physical and chemical properties of gas giants across the full temperature range spanned by hotter directly imaged exoplanets and the cooler giants of the Solar System.   
}
{{We characterize the atmosphere of Epsilon Indi Ab using infrared spectroscopic observations.}}
{We present JWST spectroscopic observations of Epsilon Indi Ab obtained with MIRI in the medium-resolution spectrograph mode (4.9--27.9~$\mu$m) and NIRSpec in the integral field unit mode (2.87--5.27~$\mu$m). The data are highly contaminated by the stellar light from the bright host star. We therefore adopted a high-pass filtering and cross-correlation approach based on \texttt{petitRADTRANS} atmospheric models to detect the planetary signal and specific molecules.}
{The cross-correlation analysis enabled us to robustly detect the planet with both instruments with a signal-to-noise ratio (S/N) up to $\mathrm{S/N}=15.0$. Using molecular mapping, we securely identified the presence of \ch{NH3} (maximum $\mathrm{S/N} = 14.4$), \ch{H2O} (maximum $\mathrm{S/N} = 11.3$), \ch{HCN} ($\mathrm{S/N} = 5.3$) and marginally detected \ch{PH3} ($\mathrm{S/N} = 4.1$), \ch{CH4} ($\mathrm{S/N} = 3.8$), \ch{C2H2} ($\mathrm{S/N} = 3.9$) and \ch{CO2} ($\mathrm{S/N} = 2.6$). In particular, Epsilon Indi Ab becomes the coldest exoplanet in which \ch{NH3} and \ch{HCN} have been identified so far. The presence of HCN likely implies intense vertical mixing.
}
{These first results confirm that infrared spectroscopy combined with cross-correlation techniques can be used to characterize the atmosphere of directly imaged exoplanets with JWST, even when the contamination by the host star is critical. 
}


   \maketitle
   \nolinenumbers

\section{Introduction}
\nolinenumbers

The study of cold (under 300 K) wide-orbit giant exoplanets represents a crucial frontier in exoplanetary science, as these objects serve as direct analogs to the gas giants of our own Solar System. It is vital to observe and characterize such objects to understand the formation, migration, and long-term atmospheric evolution of planetary systems. However, cold directly imaged planets remain exceptionally rare. Until recently, ground-based facilities were heavily limited in the infrared by the intense thermal background emitted by the sky and the telescope itself at longer wavelengths, along with, to a lesser extent, strong telluric absorption from atmospheric water vapor and \ch{CO2} and OH airglow emission lines in the near-infrared. This left the intrinsic thermal emission of these cold worlds largely inaccessible \citep{TraubOppenheimer_2010, Muller_2020}. The James Webb Space Telescope (JWST) now offers a unique opportunity to directly image and analyze the emission from this class of objects through its high infrared sensitivity and high-contrast imaging capabilities \citep{Brande_2019, Carter_2023}.

\begin{figure}[htbp]
\centering
\begin{tikzpicture}
    \node[anchor=south west, inner sep=0] (image) at (0,0) {\includegraphics[width=\linewidth]{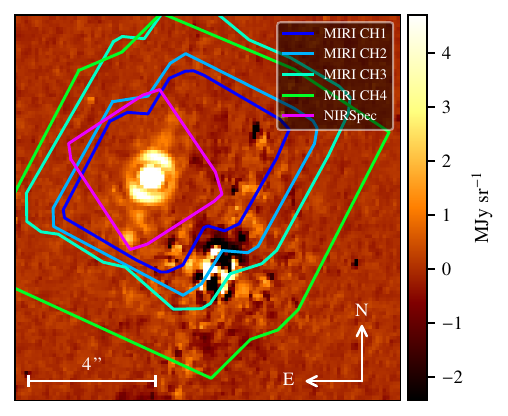}};
    
    \begin{scope}[x={(image.south east)}, y={(image.north west)}]
        \node[anchor=center] at (0.425, 0.35) {\includegraphics[width=0.09\linewidth]{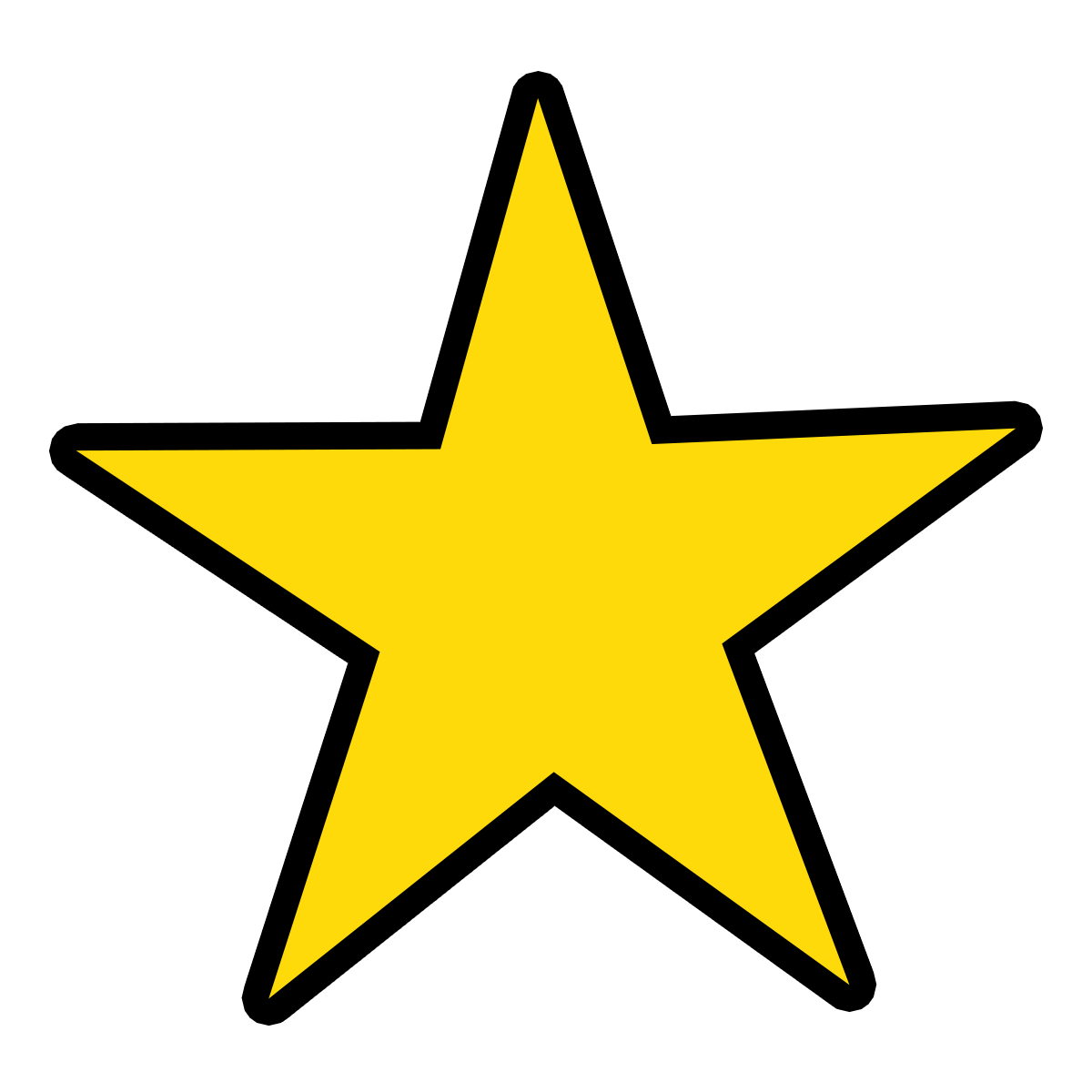}};
    \end{scope}
\end{tikzpicture}
\caption{Coronagraphic image of Eps Ind A (masked in the lower center with a star schema), obtained with the F1140C filter of JWST/MIRI by \citet{Matthews_2026}. Epsilon Indi Ab is detected as a bright point source in the upper left corner of this image. The fields of view of MIRI MRS and NIRSpec IFU presented in this paper are shown as colored footprints.}
\label{fig:corono}
\end{figure}

A few cold planetary-mass companions are currently known, such as the widely separated (130") WD 0806-661 b \citep{Luhman_2011, Luhman_2014, Lew_2026} or candidates like GJ 504 b \citep{Janson_2013}. To date, the strongest case for a true Jupiter analog is Epsilon Indi Ab \citep{Matthews_2024}. Located approximately 3.6 pc from Earth, Epsilon Indi Ab is a gas giant exoplanet orbiting the K-type star Epsilon Indi A. It was initially discovered through radial velocity measurements and astrometry \citep{Endl_2002, Zechmeister_2013, Philipot_2023, Feng_2023}, but recent direct imaging with the MIRI (Mid-InfraRed Instrument) coronograph on the JWST has provided more accurate constraints on its mass, orbit, and atmospheric properties \citep{Matthews_2024, Matthews_2026}. This observation is shown in Fig. \ref{fig:corono} and was retrieved from the MAST archive\footnote{Mikulski Archive for Space Telescopes: \url{https://mast.stsci.edu}} (Program GO 5037, PI: Elisabeth Matthews).
Epsilon Indi Ab is particularly notable for being one of the coldest directly imaged exoplanets, with an estimated effective temperature of 275 K \citep{Baudino_2019, Matthews_2024}. This is three times the effective temperature of Saturn and twice that of Jupiter, and it is close to the condensation temperature of water. Because it is far from its star (around 20 AU), the Epsilon Indi Ab equilibrium temperature is particularly low (below 100 K, estimated based on an approximate albedo), and its approximate effective temperature of 275 K is probably due to the planet's internal heat. With an estimated age of 2.2 to 5.7 billion years \citep{Feng_2019, Chen_2022} and assuming a radius of $\sim$1 R$_{Jupiter}$ (typical for planets of this age), the effective temperature of $\sim$275 K indicates a mass of $\sim$6 M$_{Jupiter}$. Because the age of the Solar System is $\sim$4.6-billion years \citep{Connelly_2012}, Epsilon Indi Ab offers a valuable analog for studying the atmospheric properties of evolved gas giants. \\

Preliminary analysis using the first MIRI coronagraphic images and archival VLT-VISIR (VLT Imager and Spectrometer for the mid-InfraRed) data allowed the extraction of three photometric measurements at 10.65, 10–12.5, and 15.50 $~\mu$m \citep{Matthews_2024}. These initial results highlighted a significant puzzle: the source remained undetected at shorter wavelengths, with VLT-NACO (Nasmyth Adaptive optics system \& COude near infrared camera) imaging setting stringent upper limits on the near-infrared flux \citep{Viswanath_2021} at 3.80 $\mu$m. Planetary atmosphere models were able to reproduce this lack of NIR emission by either invoking a high metallicity (higher than 1.0) and/or a carbon-rich environment with large abundances of methane (\ch{CH4}; \citealt{Matthews_2024}). 
More recent photometric campaigns by  \citet{Matthews_2026}
have provided deeper insights into this atmospheric mystery. These updated data have revealed tentative evidence of ammonia (\ch{NH3}) absorption and provided strong support for the presence of water-ice (\ch{H2O}) clouds as the primary source of opacity suppressing the near-infrared flux. Nevertheless, photometry alone cannot break the degeneracies between cloud properties, temperature, metallicity, radius, and chemical abundances \citep{Daemgen_2017, Inglis_2024}. High-quality spectroscopy is required to obtain a detailed view of the atmospheric composition and thermal structure and to confirm the planetary nature of the signatures. Simulations and early release science observations have already demonstrated the capabilities of the JWST to obtain direct spectroscopy with a high-signal-to-noise ratio of planetary-mass companions that are separated from their host star by just a few arcseconds \citep{Patapis_2022, Malin_2023, Miles_2023}.
Taking advantage of these capabilities, we obtained observations of Epsilon Indi Ab using  the MIRI MRS (Medium Resolution Spectrometer) IFU (Integral Field Unit) in the $5-28~\mu m$ range, together with NIRSpec (Near-Infrared Spectrograph) IFU in the $2.87 - 5.27~\mu m$ range.\\

In this paper, we present a first analysis of this spectroscopic data using the combined NIRSpec and MIRI data. The paper is organized as follows: in Sect. \ref{sec:Observations and data reduction}, we describe the observations and data reduction; in Sect. \ref{sec:Results and analysis}, we present the data, the detection of the planet, and the molecular mapping; in Sect. \ref{sec:Discussion} we discuss these results, and in Sect. \ref{sec:Conclusion} we give the main conclusions.

\section{Observations and data reduction}
\label{sec:Observations and data reduction}

The JWST spectroscopic observations of Epsilon Indi Ab with NIRSpec and MIRI are part of the JWST GO program PID 8438 (P.I. Olivier Berné)\footnote{MAST DOI: \url{https://doi.org/10.17909/m5aq-1338}}.

\vspace{-3.8ex}
\paragraph{MIRI MRS:} The MIRI MRS observations covered the wavelength range 4.9--27.9 $\mu$m by combining its four channels 1, 2, 3 and 4, each of which is further divided into three sub-channels A, B, and C (the wavelength ranges of each sub-channels are given in Table~\ref{tab:detection}), with a resolving power between 1300 (channel 4) and 3700 (channel 1). The MIRI MRS observations used the SLOWR1 readout mode to limit the data rate and a four-point dither pattern with two integrations and 58 groups per integration. The total science exposure time was 5781.36 s, and the observations were carried out on 18 October 2025. Each visit included a dedicated background observation in order to optimize the data quality given the faintness of the source (which is estimated to be between 50 and 100 MJy/sr within the maximum emission ranges). The pointing for these observations was based on approximate orbital parameters provided by E.~C.~Matthews (priv. comm.) that were derived from photometric observations.

\vspace{-3.8ex}
\paragraph{NIRSpec IFU:} The NIRSpec IFU observations, executed using the NRSIRS2RAPID readout mode with the G395H grating combined with the F290LP filter, covered the wavelength range 2.9--5.3 $\mu$m, with a resolving power of 2700. The IFU observations consisted of a four-point dither pattern over the exposures. Each exposure included one integration and 50 groups per integration. The total science exposure time was divided into two observations of 2917.776 s each, one on 18 May 2026, and one on 03 July 2026. The pointing for these observations was based on the exact position of the planet found in the MIRI observations. To do this, we first determined the exact position of the planet at the time of the MIRI MRS observations: the position of Epsilon Indi Ab was identified as the location of the maximum cross-correlation (CC) signal in the detection maps (Sect.~\ref{subsec:Cross-correlation processing}). Starting from this reference position (October 2025), we then propagated the orbit using the derived orbital elements provided by \cite{Matthews_2026} to predict the planet coordinates at the epochs of the NIRSpec observations. The maximum of the CC map in the NIRSpec cube (Sect.~\ref{subsec:Cross-correlation processing}) is compatible with the predicted coordinates within 0.1" (IFU spaxel size), equaling the JWST pointing accuracy (0.10"). We were therefore able to validate the orbital solution reported by \cite{Matthews_2026}.\\
For this study, we directly retrieved the fully reduced stage 3 data cubes from the MAST archive.

\section{Results and analysis}
\label{sec:Results and analysis}

\begin{figure*}
    \centering

    \includegraphics[width=0.33\textwidth]{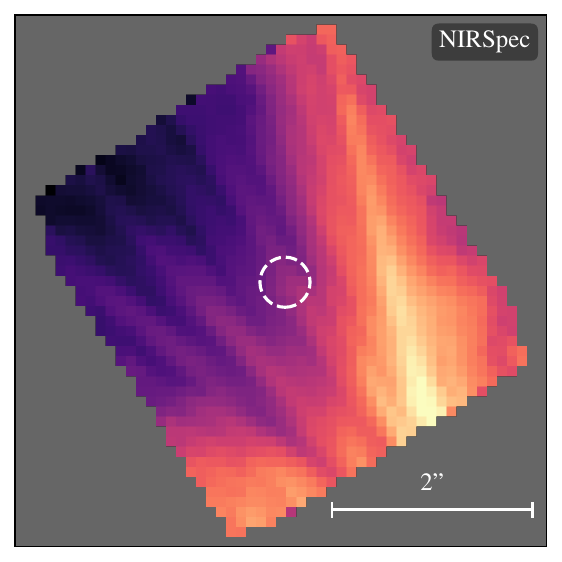}
    \hfill
    \includegraphics[width=0.33\textwidth]{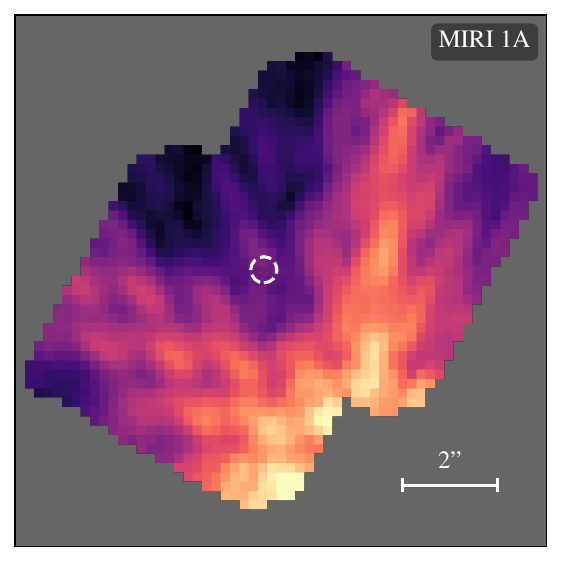}
    \hfill
    \includegraphics[width=0.33\textwidth]{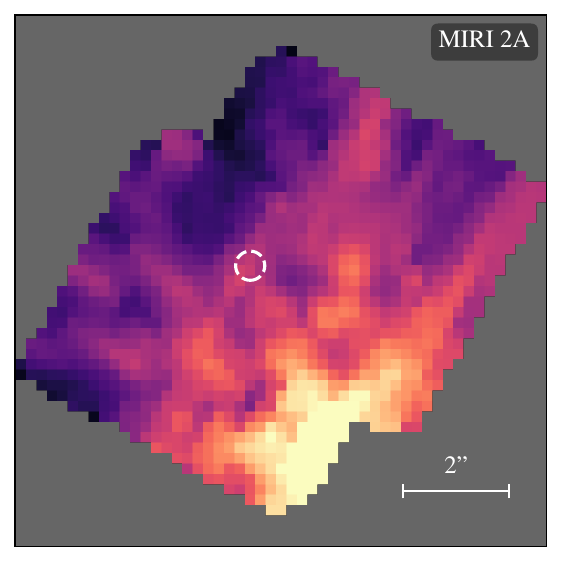}
    
    \vspace{0.5em}

    \makebox[0.094\textwidth]{} 
    \includegraphics[width=0.33\textwidth]{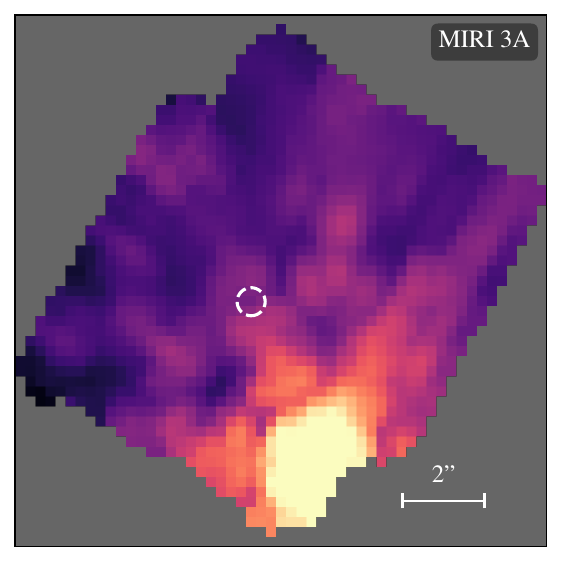}
    \hfill
    \includegraphics[width=0.33\textwidth]{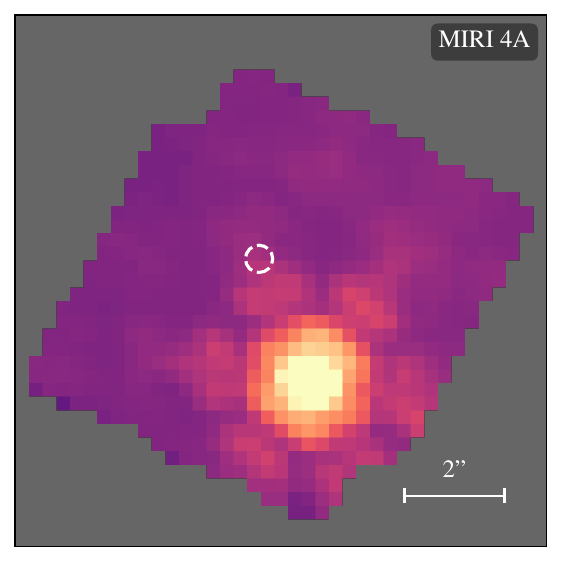}
    \hfill
    \includegraphics[width=0.13\textwidth]{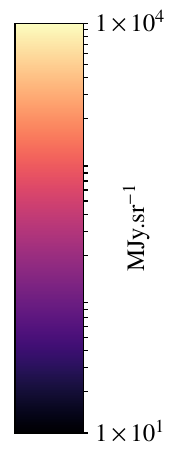}
    \makebox[0.094\textwidth]{} 
    
    \caption{Surface brightness maps from the NIRSpec and MIRI cubes, corresponding to the average value across the spectral channels of each cube (the surface brightnesses of the channels were summed and divided by the total number of channels). For NIRSpec, the spectral channels between 3.97 and 4.22\,$\mu$m are masked to exclude the detector gap between the two camera arrays. The dotted white circle indicates the predicted planet location using orbital parameters. North is up in all images.}
    \label{fig:flux_maps}
\end{figure*}

\subsection{Data}
\label{subsec:Data}

The data used for this study are presented in Fig. \ref{fig:flux_maps}.  The surface brightness maps are dominated by the stellar point-spread function (PSF). The spectral cubes are also dominated by the PSF at all wavelengths. There are several methods to address this contamination effect: performing PSF subtraction using models, using reference-star differential imaging (RDI), using angular differential imaging (ADI), or using a cross-correlation analysis to detect the spectroscopic signal of the planet without direct subtraction of the PSF.

We first attempted to subtract the PSF of the star Epsilon Indi A from our data in order to recover the planetary signal. 
Because of the large angular distance ($\sim$ 6'') between the planet and host star, the two are not observed simultaneously with NIRSPec and MIRI up to channel 2 (Fig.~\ref{fig:flux_maps}). This makes angular differential imaging (ADI) challenging, and in addition, the MIRI observations were obtained in a single position angle, which precludes ADI. The NIRSpec observations were obtained at two epochs, but the variation in the angle between the two visits (38.63 degrees) was such that the PSF had rotated too much to allow subtraction at the position of the planet. The portion of the stellar PSF at the planet location in the first observation lies outside the field of view in the second observation.

We attempted to model the stellar PSF using the Python module \texttt{STPSF}\footnote{\url{https://stpsf.readthedocs.io}} \citep{Perrin_2012, Perrin_2014}. Since the planet is several arcsecs away from its star and the field of view is centered on the planet, the center of the stellar PSF is located outside the field of view in NIRSpec data and MIRI channels 1 and 2, which prevented us from precisely centering and scaling the PSF for a subtraction for these channels. We therefore only tried the subtraction on MIRI channels 3 and 4. As \texttt{STPSF} is designed for photometry, we reconstructed a PSF for each wavelength plane of a cube (i.e., a sub-channel), which amounted to 500 to 1,000 PSFs per cube. We then subtracted it plane by plane, giving us a subtraction of the PSF on the whole cube. 
We obtained residuals that were too large compared to the expected signal from the planet, however. According to photometric measurements of \citet{Matthews_2024}, Epsilon Indi Ab is expected to produce a maximum surface brightness of 50 to 100 MJy/sr in the MIR pixels, whereas we found residuals exceeding 100 MJy/sr. We therefore tried to use RDI, which involves modeling the PSF using a principal component analysis PCA) based on various reference objects that were observed during cycles 1 and 2, such as asteroids or isolated stars. Nevertheless, the residuals were once again too large (exceeding 100 MJy/s) to allow us to conclude that the planet had been detected.\\

\subsection{Cross-correlation processing}
\label{subsec:Cross-correlation processing}

Considering the difficulty to perform PSF subtraction on our data, we turned to methods relying on cross-correlation using a model spectrum of the planet. This approach is expected to be particularly effective for cold objects \citep{Patapis_2022}, such as Epsilon Indi Ab, that exhibit low contrast. Cross-correlation is a statistical operation that measures the similarity between two signals as a function of a relative shift applied to one of them, and it has been proposed to detect the presence of a known template within a noisy dataset for exoplanet detection \citep{Sparks_2002}. This technique has emerged as a powerful approach in the presence of stellar PSF contamination and instrumental noise in IFU data (see, e.g., \citealt{Hoeijmakers_2018, Petrus_2021, Patapis_2022}). 
This approach is particularly valuable in cases like ours, where subtraction of the stellar PSF down to the planetary contrast level is difficult. We applied the method described in \cite{Patapis_2022}, which relies on \texttt{petitRADTRANS}\footnote{\url{https://petitradtrans.readthedocs.io}} \citep{Molliere_2019}, to perform the cross-correlation detection\footnote{We note that we also successfully applied this method using \texttt{Sonora Elf Owl} models \citep{Mukherjee_2024}, but because these are limited to a maximum wavelength of 15 $\mu$m, this precludes an analysis of the longer wavelengths of MIRI-MRS. Consequently, we focused on the results obtained with \texttt{petitRADTRANS}.}. 

The cross-correlation coefficient between an observed spectrum $s_\mathrm{obs}(\lambda)$ and a theoretical template $s_\mathrm{theo}(\lambda)$ is computed as \citep{Zucker_2003} 

\begin{equation}
    \hfill
    \mathrm{CC} = \frac{1}{N} \sum_{i=1}^{N} 
    \frac{\tilde{s}_\mathrm{obs}(\lambda_i)}{\sigma_\mathrm{obs}} \cdot 
    \frac{\tilde{s}_\mathrm{theo}(\lambda_i)}{\sigma_\mathrm{theo}},
    \hfill
    \label{eq:cc}
\end{equation}

\noindent where N is the number of spectral bins over which the sum is computed, $\tilde{s} = s - \langle s \rangle$ denotes the mean-subtracted signal, and $\sigma$ its RMS amplitude, such that $\mathrm{CC} \in [-1, 1]$. This operation was applied independently to the spectrum extracted from each spatial pixel (spaxel) of the IFU cube after binning the flux within a circular aperture of radius $r_\mathrm{bin}$ pixels. This yielded a two-dimensional cross-correlation map $\mathrm{CC}(x, y)$ across the full field of view. The S/N at a given spaxel was then defined as

\begin{equation}
    \hfill
    \mathrm{S/N}(x, y) = \frac{\mathrm{CC}(x, y)}{\sigma_\mathrm{local}},
    \label{eq:snr}
    \hfill
\end{equation}

\noindent where $\sigma_\mathrm{local}$ is the standard deviation of the CC map 
estimated within an annular region $[R_\mathrm{inner},\, R_\mathrm{outer}]$ 
centered on the CC peak, and serves as a local estimate of the noise level 
\citep{Mawet_2014}.
We considered an annulus with an inner radius of 12 pixels and an outer radius of 18 pixels. These values were chosen to minimize the residues of the planet and the star PSFs in the annulus.

The theoretical models were computed with \texttt{petitRADTRANS}, assuming H$_2$ and He as the main continuum opacity sources, including collision-induced absorption (CIA H$_2$-H$_2$ and H$_2$-He) and Rayleigh scattering, with mass fractions of 0.74 and 0.24, respectively. This is consistent with solar abundances \citep{Asplund_2021} and commonly adopted as default values in atmospheric models of giant exoplanets such as \texttt{petitRADTRANS}. We also included CH$_4$, H$_2$O, NH$_3$, CO, and CO$_2$, whose abundances were specified through their mass-mixing ratios (MMRs), corresponding to their mass fractions. We adopted typical MMR values, such as $10^{-3}$ for CH$_4$ and $10^{-5}$ for CO$_2$, that are representative of similar objects \citep{Molliere_2019, Wang_2023, Miles_2023, Lew_2026}. The atmospheric mean molecular weight was fixed at 2.33, consistent with the assumed \ch{H2}/\ch{He} composition. Finally, the atmospheric temperature-pressure profile was computed using the analytical parameterization of \citep{Guillot_2010}, based on infrared and visible opacities, the internal temperature (here, $\sim$275K, the estimated effective temperature), and the equilibrium temperature ($\sim$50 K) of the planet\footnote{We estimated the equilibrium temperature assuming a Bond albedo of A$\sim$0.5, which is approximately the Jupiter Bond albedo. \citep{Li_2018}.}.

The cross-correlation technique requires application of a high-pass filter to the data before cross-correlation in order to remove contamination from the stellar signal (low-frequency component) while preserving the high-frequency molecular features of the planetary spectrum \citep{Cabot_2018, Patapis_2022}. Examples of the power spectrum as a function of frequency are presented in Appendix~\ref{sec:Power spectrum}. The portion of the spectra beyond a spatial cutoff frequency of approximately 10 $\mu$m$^{-1}$ is generally dominated by noise and the planetary molecular signatures, whereas the stellar continuum is contained in the lower-frequency components. The choice of the spatial cutoff frequency $f_c$ involves a trade-off: a value that is too low retains residual stellar continuum, which increases the background noise, whereas a value that is too high removes genuine planetary spectral features, thereby reducing the cross-correlation signal \citep{Hoeijmakers_2018, Petit_2018, Bidot_2024}. The spectral cube and the theoretical model were filtered independently using a fourth-order Butterworth high-pass filter applied pixel by pixel along the spectral axis.\\
Since we have prior knowledge of the approximate sky coordinates of Epsilon Indi Ab (used for the telescope pointing), but cannot be certain that they are exact\footnote{The uncertainty arises from the use of preliminary orbital parameters for the MIRI observations and from the absence of target acquisition for NIRSpec, resulting in a possible pointing error of $0.1''$.}, we first applied the same cutoff frequency, $f_c = 10\,\mu\mathrm{m}^{-1}$, to each data cube (the NIRSpec cube and the 12 MIRI sub-channel cubes). At this cutoff frequency, the planet is already detected in several cubes using cross-correlation. We used these initial detections to determine the exact position of the planet.
The determination of the cutoff frequency is described in detail in Appendix~\ref{sec:S/N and spatial cutoff frequency}. Briefly, we injected eight simulated planets to the data cubes and calculated the optimal cutoff frequency defined as the frequency that maximizes the recovered S/N (derived with the method described in Sect. \ref{subsec:Cross-correlation processing} and Eq. \ref{eq:snr}) for the simulated planet.  We restricted the optimization to the physically plausible range $f_c \in [0.5,\,20]\,\mu\mathrm{m}^{-1}$ \citep{Bidot_2024}. The average of the optimal frequencies then yields the cutoff frequency $f_c^{\star}$ that we applied to our data.

\begin{figure*}
    \centering

    \includegraphics[width=0.24\textwidth]{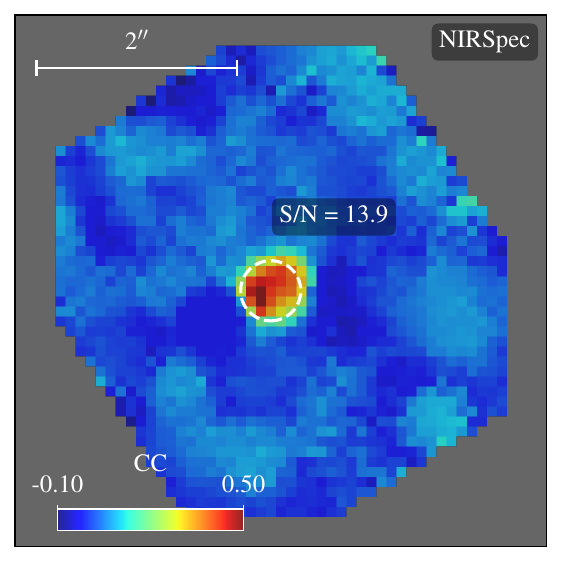}
    \hfill
    \includegraphics[width=0.24\textwidth]{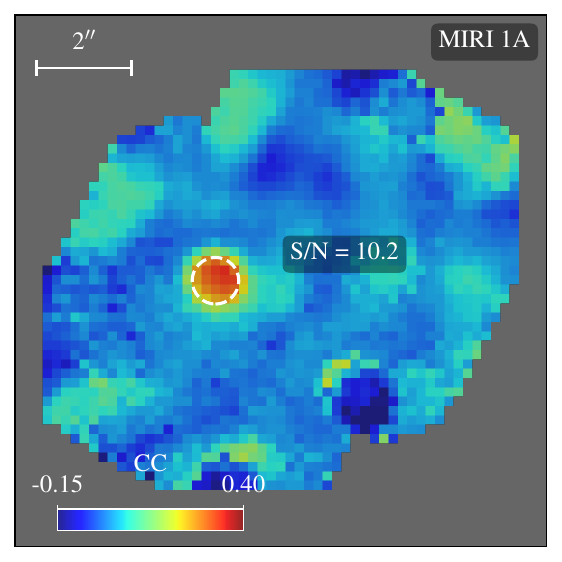}
    \hfill
    \includegraphics[width=0.24\textwidth]{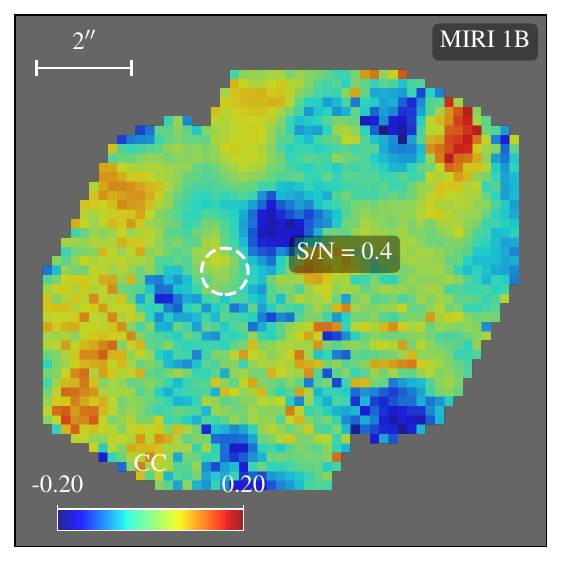}
    \hfill
    \includegraphics[width=0.24\textwidth]{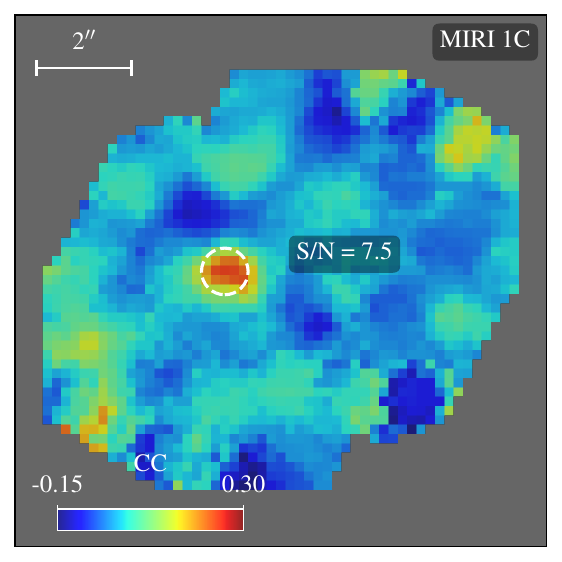}

    \vspace{0.5em}
        
    \includegraphics[width=0.24\textwidth]{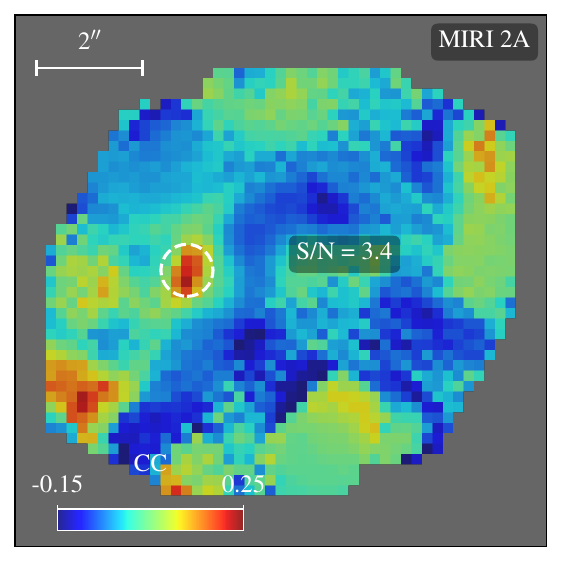}
    \hfill
    \includegraphics[width=0.24\textwidth]{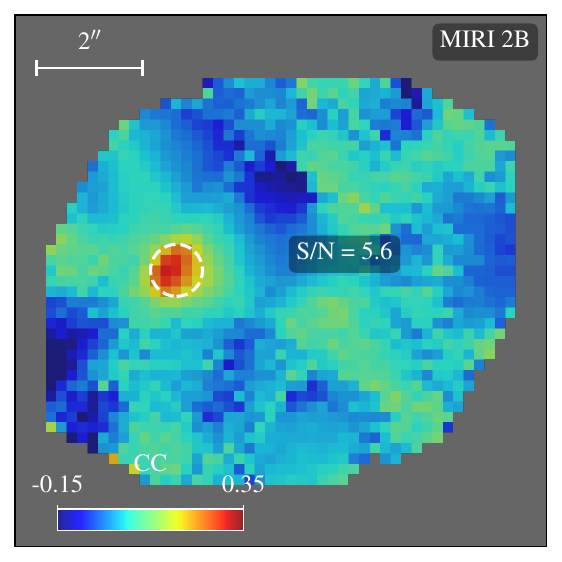}
    \hfill
    \includegraphics[width=0.24\textwidth]{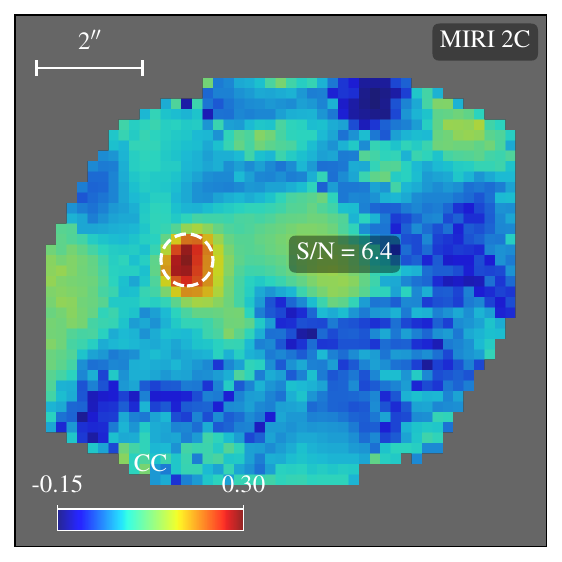}
    \hfill
    \includegraphics[width=0.24\textwidth]{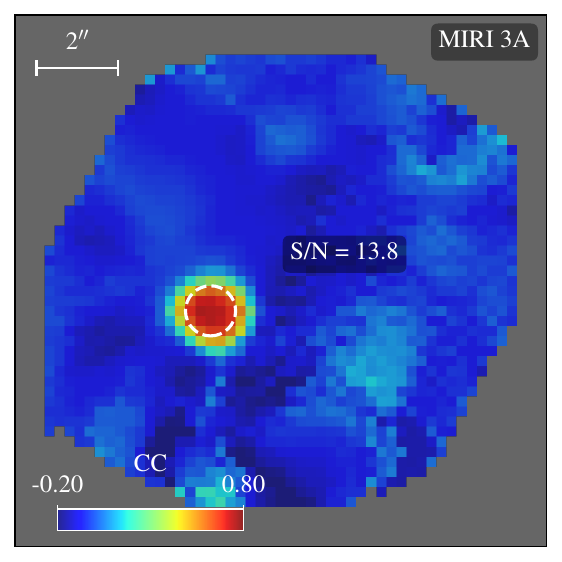}
    
    \vspace{0.5em}

    \makebox[0.24\textwidth]{}
    \hfill
    \includegraphics[width=0.24\textwidth]{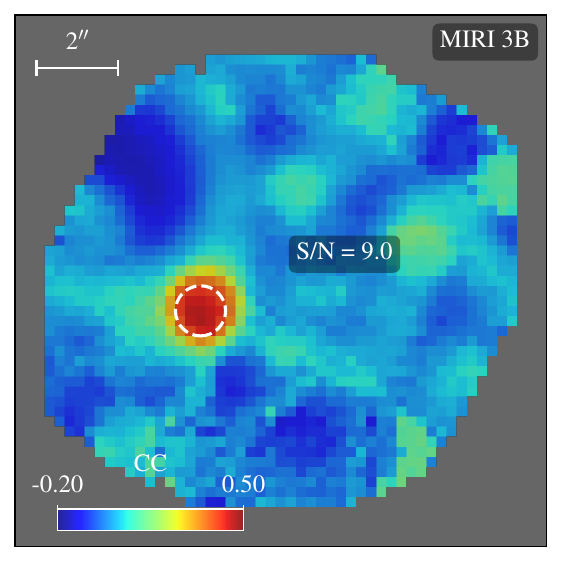}
    \hfill
    \includegraphics[width=0.24\textwidth]{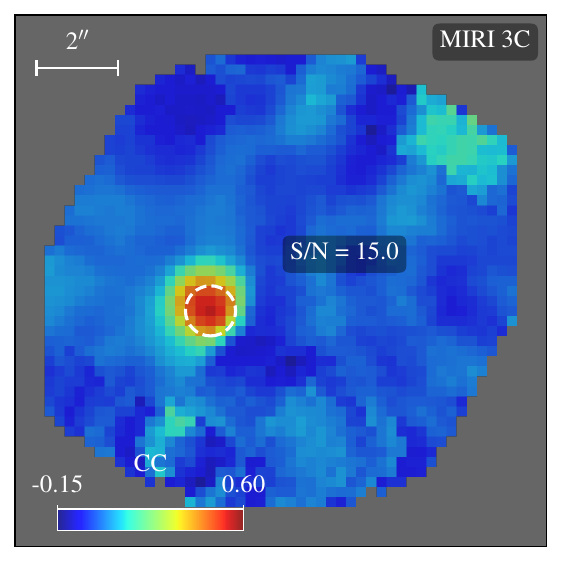}
    \hfill
    \makebox[0.24\textwidth]{}
    
    \caption{CC maps for Epsilon Indi Ab, with NIRSpec IFU and MIRI MRS sub-channels. We excluded the CC maps of MIRI 4A, 4B, and 4C because their sensibility is very low. North is up in all images.
    }
    \label{fig:cc_maps_total}
\end{figure*}

Table~\ref{tab:detection} summarizes the S/N values determined for each cube, along with the associated spectral range and the cutoff frequency. The cutoff frequencies are available in pixel$^{-1}$ (spectral pixel) in Appendix \ref{sec:S/N and spatial cutoff frequency}. The CC maps we derived are also shown in Fig.~\ref{fig:cc_maps_total}.
\begin{table}[ht!]
\caption{Detection of Epsilon Indi Ab using cross-correlation.}
\centering
\begin{tabular}{lccc}
\hline\hline
Instrument & Spectral range [$\mu$m] & S/N & $f_c^{\star}$ [$\mu\mathrm{m}^{-1}$]\\
\hline
NIRSpec -  G395H       &2.87--5.27    &13.9     &15.24\\
MIRI - 1A              &4.90--5.74    &10.2     &15.19\\
MIRI - 1B              &5.66--6.63    &0.4      &8.27\\
MIRI - 1C              &6.53--7.65    &7.5      &7.00\\
MIRI - 2A              &7.51--8.77    &3.4      &3.80\\
MIRI - 2B              &8.67--10.13   &5.6      &2.28\\
MIRI - 2C              &10.01--11.70  &6.4      &3.42\\
MIRI - 3A              &11.55--13.47  &13.8     &3.61\\
MIRI - 3B              &13.34--15.57  &9.0      &12.48\\
MIRI - 3C              &15.41--17.98  &15.0     &12.22\\
MIRI - 4A              &17.70--20.95  &3.5      &7.49\\
MIRI - 4B              &20.69--24.48  &0.3      &8.66\\
MIRI - 4C              &24.40--27.90  &0.0      &10.42\\
\hline
\end{tabular}

\label{tab:detection}
\end{table}
 The planet is clearly detected in the NIRSpec data (S/N of 13.9) and in the MIRI sub-channels 1A (10.2), 3A (13.8), 3B (9.0), and 3C (15.0). In sub-channels 1C, 2B, and 2C, the planet is also detected with a weaker signal (S/N of 7.5, 5.6, and 6.4, respectively). In sub-channel 1B, the non-detection (S/N of 0.4) is associated with spectral regions of strong atmospheric absorptions. Finally, channel 4 is affected by the decline in MIRI sensitivity at the longest wavelengths \citep{Glasse_2015}. Appendix~\ref{sec:S/N and spectrum} explains the link between spectral emission and S/N in more detail.

\subsection{Molecular mapping}
\label{subsec:Molecular mapping}

Molecular mapping \citep{Hoeijmakers_2018, Petit_2018} is a spectroscopic technique that is used in the direct imaging of exoplanets to detect and spatially localize molecular species in a planetary atmosphere. It relies on cross-correlating the spectrum extracted at each spaxel of an integral field spectrograph datacube with a synthetic template spectrum.

Molecular mapping extends the previously described procedure by repeating it for different target molecules (\ch{CH4}, \ch{NH3}, \ch{H2O}, \ch{CO}, and \ch{CO2}), with each synthetic template including only one species at a time in the modeled atmosphere. 
Thus, the resulting S/N map exclusively reflects the contribution of this species to the planetary signal: a molecule that is absent from the atmosphere will not produce any coherent structure in the correlation map, while a molecule that is present will generate a peak located at the position of the source.
Molecular mapping has been applied successfully to synthetic JWST data \citep{Patapis_2022, Malin_2023}, and more recently, to real data with the observation of the planetary-mass companion HS 1256 b \citep{Malin_2026}.

\begin{figure*}[h!]
    \centering

    \includegraphics[width=0.24\textwidth]{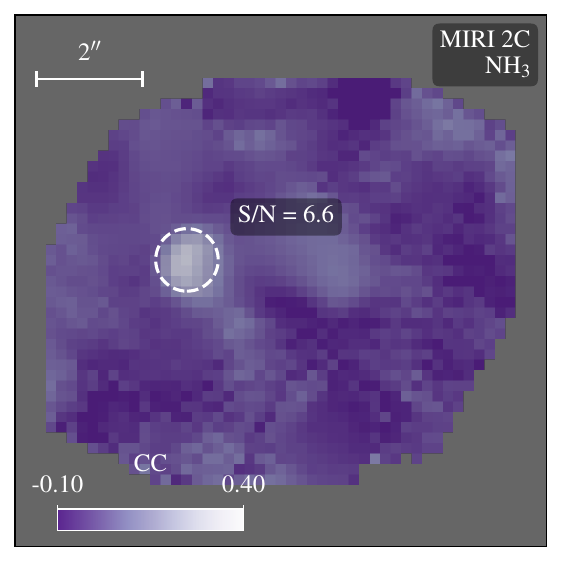}
    \hfill
    \includegraphics[width=0.24\textwidth]{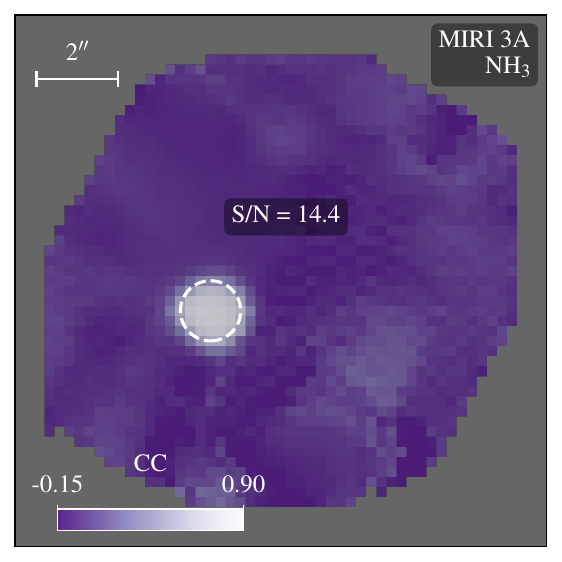}
    \hfill
    \includegraphics[width=0.24\textwidth]{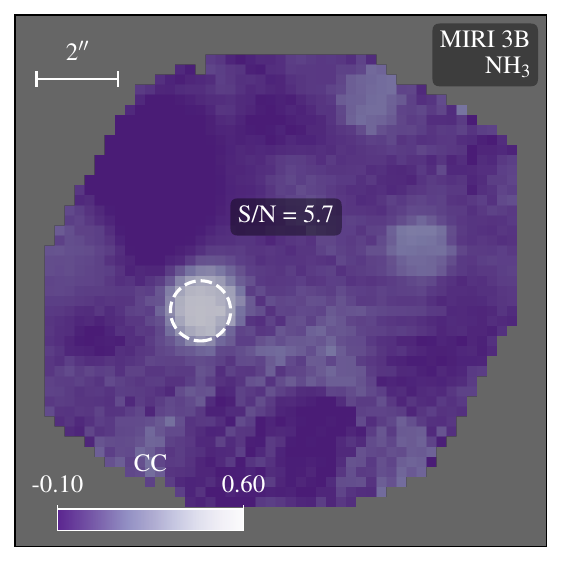}
    \hfill
    \includegraphics[width=0.24\textwidth]{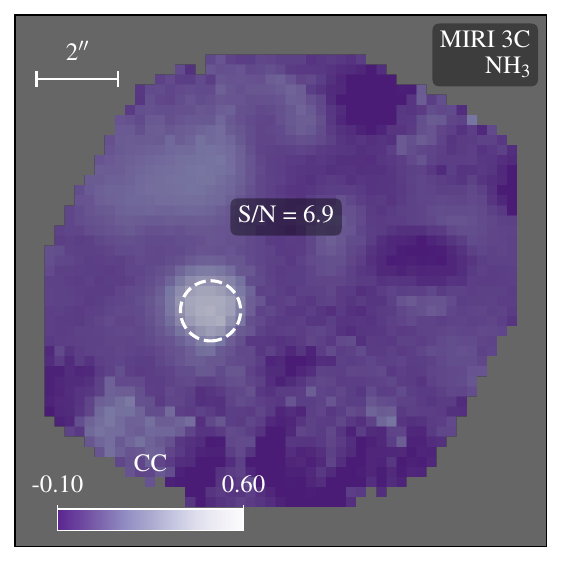}
    
    \vspace{0.5em}

    \includegraphics[width=0.24\textwidth]{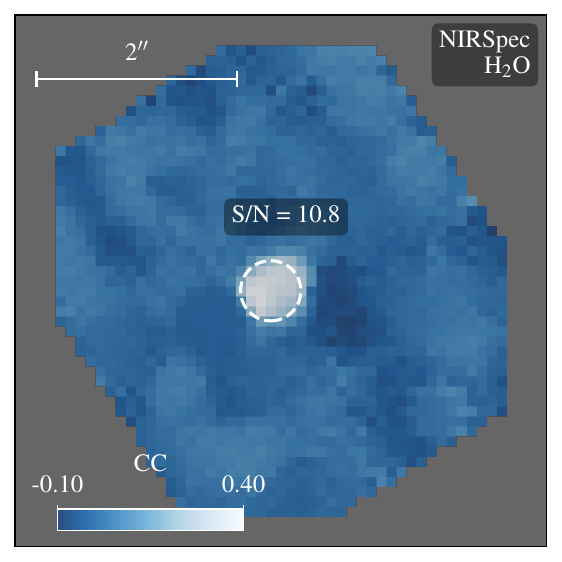}
    \hfill
    \includegraphics[width=0.24\textwidth]{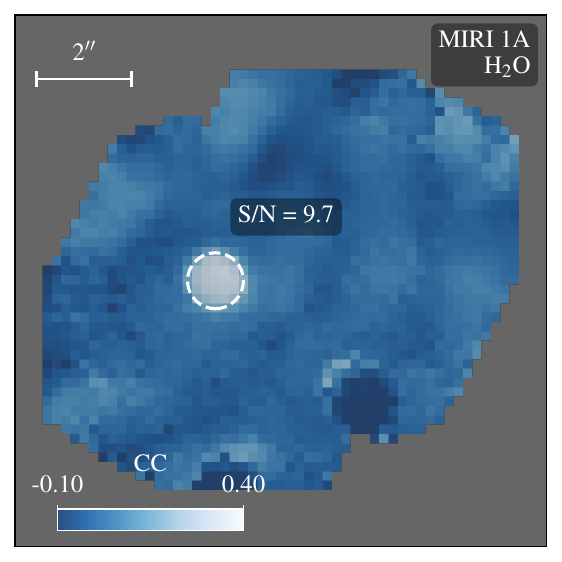}
    \hfill
    \includegraphics[width=0.24\textwidth]{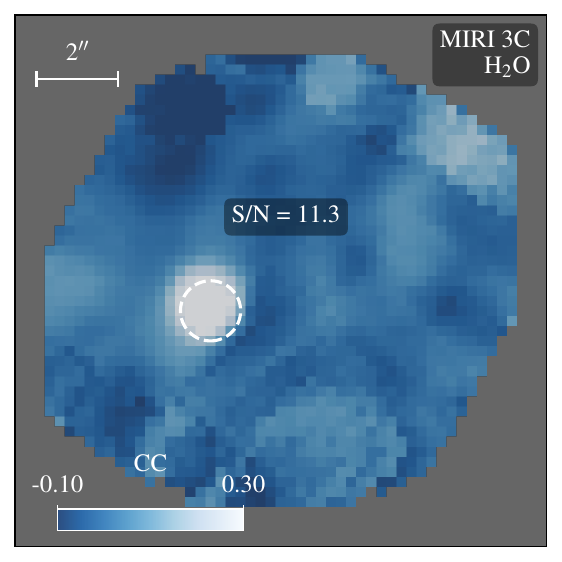}
    \hfill
    \includegraphics[width=0.24\textwidth]{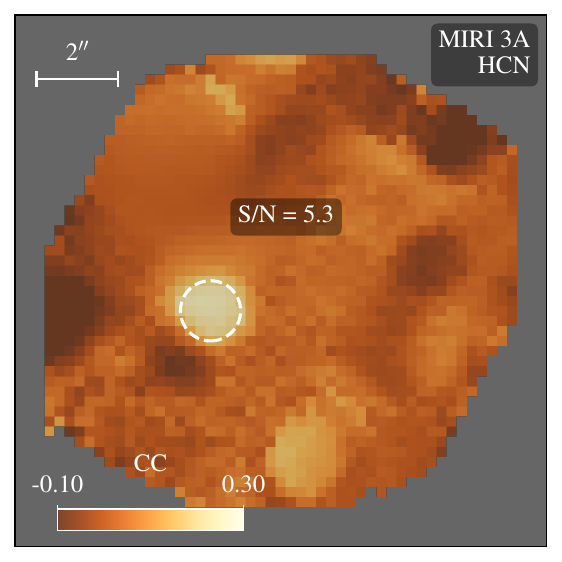}

    \vspace{0.5em}

    \includegraphics[width=0.24\textwidth]{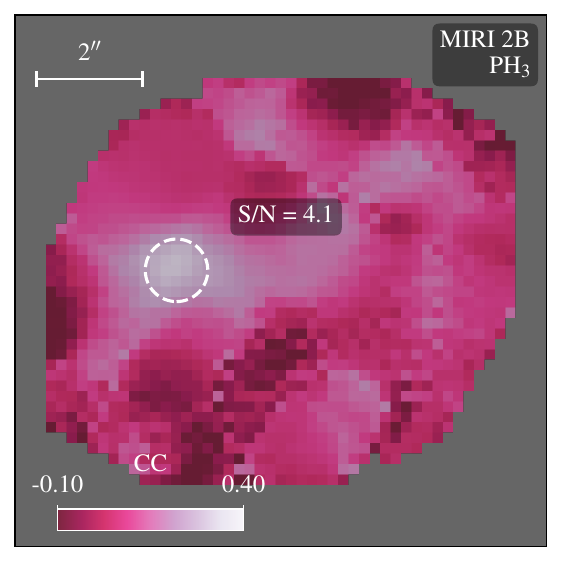}
    \hfill
    \includegraphics[width=0.24\textwidth]{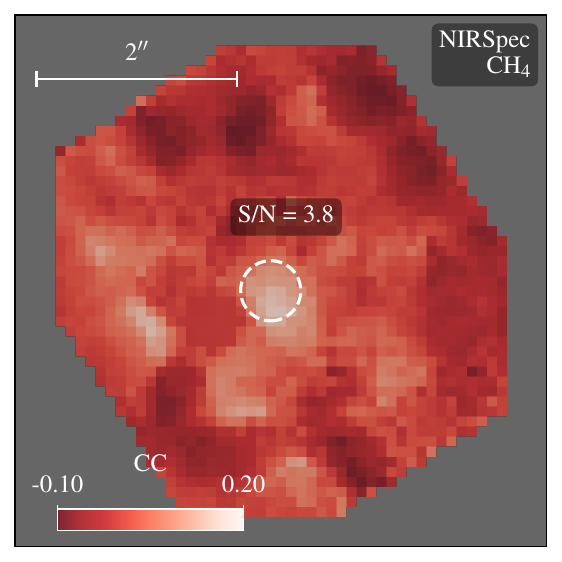}
    \hfill
    \includegraphics[width=0.24\textwidth]{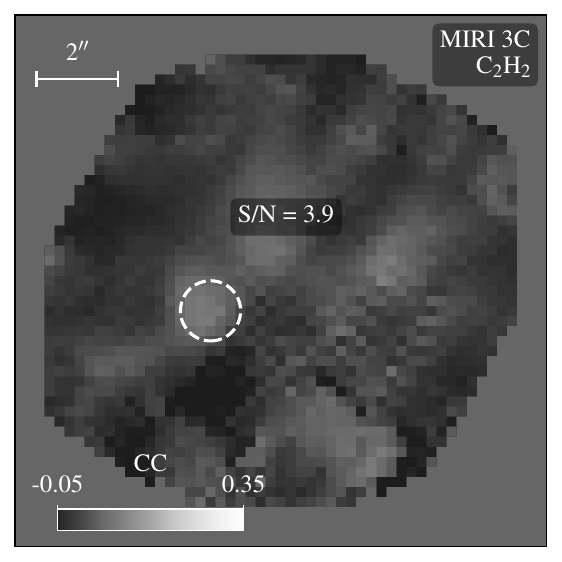}
    \hfill
    \includegraphics[width=0.24\textwidth]{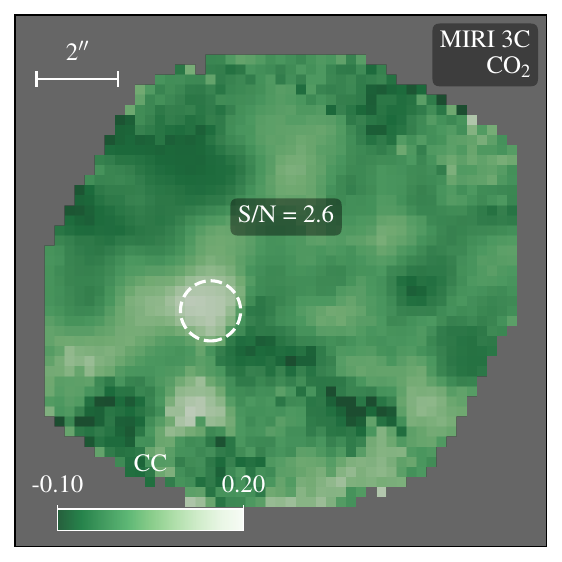}
    
    \caption{Molecular mapping results for Epsilon Indi Ab. Blue shows H$_2$O, purple shows NH$_3$, orange shows HCN, red shows CH$_4$, pink shows PH$_3$, black shows C$_2$H$_2$, and green shows CO$_2$. North is up in all images.} 
    
    \label{fig:cc_maps}
\end{figure*}

For each target species, the template was calculated with \texttt{petitRADTRANS} assuming a uniform MMR of $10^{-3}$ , a representative value producing detectable unsaturated spectral lines \citep{Patapis_2022, Malin_2023}.
The maps were then obtained using the same procedure as described in Sect.~\ref{subsec:Cross-correlation processing}. The results are presented in Fig. \ref{fig:cc_maps}.
We detected NH$_3$ in MIRI 2C, 3A, 3B, and 3C with an S/N as high as 14.4 for MIRI 3A.
We also detected H$_2$O vapor in NIRSpec, MIRI 1A, and MIRI 3C with an S/N up to 11.3 and \ch{HCN} in MIRI 3A with an S/N of 5.3. 
We note that the highest S/N values obtained in Fig. \ref{fig:cc_maps_total} are likely related to the detections of NH$_3$ adn \ch{H2O}.
A few other species were detected, but with a fainter signal: \ch{PH3} (S/N = 4.1), \ch{CH4} (S/N = 3.8), and\ch{C2H2} (S/N = 3.9). We also show an attempt to detect CO$_2$ with an S/N of 2.6 in MIRI 3C that needs to be confirmed.

\section{Discussion}
\label{sec:Discussion}

\citet{Matthews_2024} initially proposed a combination of a metal-rich and high C/O ratio 
atmosphere to explain the non-detection of the planet’s near-IR (3--5\,$\mu$m) flux. 
\citet{Matthews_2026} later measured an F1065C$-$F1140C color of $0.88 \pm 0.08$~mag, which strongly suggested the presence of a weak ammonia absorption feature. Combined with the near-infrared non-detections, this is suggestive of a cold atmosphere with water-ice clouds. 
Our data provide a direct spectroscopic confirmation of this detection through the cross-correlation analysis, making it the first spectroscopic confirmation of NH$_3$ for this planet. Additional NH$_3$ detections have been reported in similar objects, including a multiband photometric detection for GJ~504\,b \citep{Malin_2025} and a spectroscopic detection in the brown dwarf WISE~0855 \citep{Kuhnle_2025}, which exhibits a comparable temperature atmosphere (of $\sim$285K). However, this is the first NH$_3$ detection from a planet with a temperature lower than 300K.

We also present the first \ch{HCN} detection from an object under 500K. It has so far only been detected in the atmosphere of a T8.5+T9 brown dwarf binary ($\sim$500K and $\sim$600K; \citealp{Matthews_2025}) and in exoplanets with effective temperatures higher than 1000K (e.g., \citealt{Oppenheimer_2013, Giacobbe_2021}). It is rare to find \ch{HCN} in the atmosphere of such a cold planet because \ch{HCN} typically forms under high-temperature and high-pressure conditions from \ch{NH3} and \ch{CH4} \citep{Line_2011}. There are two main hypotheses to explain the formation of this molecule here: either \ch{HCN} was synthesized photochemically in the upper layers of the atmosphere of Epsilon Indi Ab, or it was formed deep below at high temperatures and was then transported to the surface through intense vertical mixing \citep{Tsai_2021, Matthews_2025}.
Since Epsilon Indi Ab is particularly far from its K-type star ($\sim 21$ AU), it is unlikely that photolysis alone can account for the formation of \ch{HCN}. Hence, it is likely that vertical mixing is efficient and that the HCN brought at high altitudes persists long enough to be detected (quenching).
Exo-REM models \citep{Baudino_2024, Radcliffe_2026} predict the presence of HCN due to chemical disequilibrium with a mixing ratio of approximately $10^{-8}$ mol/mol for an object at 300 K, log$(g) = 4$, $C/O = 0.55$, and solar metallicity. However, the required vertical mixing can only be estimated if the abundance values are known \citep{Zahnle_2014}, which is not the case with the molecular mapping technique.

These JWST detections, together with those of H$_2$O and lower-significance detections of CH$_4$, and CO$_2$ using molecular mapping, can provide further observational constraints on the atmospheric composition using a dedicated retrieval framework with Bayesian inference \citep{Brogi_2019, Gibson_2020, Lew_2026}.
It will be particularly useful in this framework to use the latest generation of atmospheric models that include patchy clouds 
(e.g., with the Exo-REM model; \citealp{Radcliffe_2026}). 
This will allow us to derive the needed molecular abundances to place constraints on the various atmospheric scenarii for Epsilon Indi Ab: 
sub-solar metallically, nitrogen depletion, and/or thick water-ice clouds, and so on. 
This effort will be the subject of a subsequent paper.

\section{Conclusion}
\label{sec:Conclusion}

Epsilon Indi Ab is the coldest directly imaged planet. Our JWST MIRI and NIRSpec spectroscopic observations provide a first glimpse into the composition of the planet's atmosphere. Our key findings are listed below.
\begin{enumerate}
    \item The stellar PSF completely dominates the data, making it very difficult to detect the planet directly, even by subtracting the PSF.
    \item Using the radiative transfer code \texttt{petitRADTRANS}, we were able to detect the planet via cross-correlation in the NIRSpec data and in several MIRI sub-channels. The non-detections correspond to strong absorption bands in the planet’s atmosphere or to losses in the instrument sensitivity (channel 4). 
    \item By using the same technique, this time with a single molecule in the model, we were able to generate molecular mapping images. We detected \ch{NH3} in the planet's atmosphere with an S/N up to 14.4, confirming the preliminary findings from photometric measurements \citep{Matthews_2026}. We also detected \ch{HCN} with an S/N of 5.3, which, given the planet's temperature of 275 K, indicates strong vertical mixing in the atmosphere. We finally detected H$_2$O with an S/N of 11.3 and \ch{PH3}, CH$_4$, \ch{C2H2}, and \ch{CO2} with a weaker S/N at the planet location.
\end{enumerate}

These results demonstrate that cross-correlation spectroscopy and molecular mapping are effective detection and characterization tools, even with a small angular separation, high contrast, or limited data quality. This is particularly relevant for JWST IFU instruments, which lack coronagraphic capabilities but can nonetheless recover planetary signals well below the stellar residuals through spectral correlation with atmospheric templates. As more directly imaged companions are observed with JWST, and in the future, with ground-based extremely large telescopes, this method offers a viable path to the atmospheric characterization for targets that would otherwise remain beyond our reach.

\begin{acknowledgements}
      This work is based on observations made with the NASA/ESA/CSA \textit{James Webb} Space Telescope. The data were obtained from the Mikulski Archive for Space Telescopes at the Space Telescope Science Institute, which is operated by the Association of Universities for Research in Astronomy, Inc., under NASA contract NAS5-03127 for JWST. These observations are associated with program \#8438. We thank E. Matthews for providing orbital parameters of Epsilon Indi Ab which allowed optimal pointing parameters for the JWST observations presented in this paper. 
      We also thank B. Charnay and A. Radcliffe for their predictions on HCN formation using Exo-REM models.
      OB, IS are funded by the Centre National d'Études Spatiales (CNES) through the APR program.

\end{acknowledgements}

\bibliographystyle{aa} 
\bibliography{bibliography.bib}

\begin{appendix}
\nolinenumbers

\section{Power spectrum}
\label{sec:Power spectrum}

\begin{figure}[h!]
    \centering
    \includegraphics[width=\linewidth]{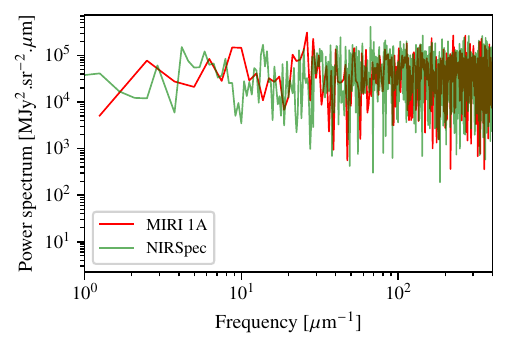}
    \caption{Power spectrum of the spectra selected at the location of Epsilon Indi Ab for MIRI sub-channel 1A and NIRSpec.}
    \label{fig:power_spectrum}
\end{figure}

Figure~\ref{fig:power_spectrum} shows the power spectrum, as a function of spatial frequency, of the spectra extracted at the location of Epsilon Indi Ab for the MIRI sub-channel 1A and NIRSpec data. In general, the power of such spectra is expected to be concentrated at low frequencies, corresponding to the stellar continuum and the smooth instrumental response, while the higher-frequency components trace the narrower molecular absorption features together with the noise \citep{Cabot_2018, Patapis_2022}. This frequency separation is the basis for the high-pass filtering strategy adopted in Sect.~\ref{subsec:Cross-correlation processing}: applying a cutoff frequency $f_c$ removes the low-frequency stellar and continuum contribution while preserving the higher-frequency planetary signal used for cross-correlation. The optimal cutoff frequency $f_c^{\star}$ (reported in Table~\ref{tab:detection}) is determined independently for each cube rather than fixed a priori, since the distribution of power across frequencies depends on the spectral resolution and sampling of each instrument and sub-channel. A single cutoff value applied uniformly to all cubes would not account for these differences, and could either leave residual stellar continuum or remove genuine planetary features depending on the data set \citep{Hoeijmakers_2018, Petit_2018, Bidot_2024}.

\section{Choice of the spatial cutoff frequency}
\label{sec:S/N and spatial cutoff frequency}

To determine which cutoff frequency to apply to each of our data subsets, we have several options available. One method involves plotting the S/N as a function of the cutoff frequency directly at the planet’s actual position. We must then select the cutoff frequency that maximizes the S/N. While this method is reliable for high S/Ns (> 5, \citealp{Mawet_2014}), it may be biased by noise randomly localized near the planet for lower S/Ns. For this reason, we use a different method: instead of considering the S/N of Epsilon Indi Ab, we insert fake planets into the data and calculate the cutoff frequency that maximizes the S/N for each of these fake planets.

To obtain credible and detectable analogs, we use the same spectrum used in cross-correlation—which was obtained using \texttt{petitRADTRANS} and convolve it with the instrumental PSF. The fake planets are then placed on a circle centered on the planet’s actual position. To ensure that the flux from Epsilon Indi Ab does not affect the S/N calculation of the fake planets, we choose a radius for this circle that is greater than three times the FWHM (Full Width at Half Maximum) of the PSF for each cube. For MIRI MRS, this quantity is calculated using the empirical linear relationship between the FWHM and the wavelength given by \citet{Law_2023}: $\rm FWHM [arcsec] = 0.033 (\lambda/\mu m) + 0.106$ arcsec. For NIRSpec IFU, this quantity is calculated using the empirical relationship given by \citet{DEugenio_2024}, which describes the major axis of the PSF: \(\mathrm{FWHM}_{\rm major}\,[\mathrm{arcsec}] = 0.12 + 1.9\,(\lambda/\mu\mathrm{m})\,\exp[-24.4/(\lambda/\mu\mathrm{m})]\). In both cases, we calculate the FWHM corresponding to the cube’s maximum wavelength in order to obtain an upper bound on the FWHM. Then to determine the cutoff frequency that maximizes the S/N for one fake planet, the entire cube is filtered and the CC map is computed over the full field of view for each candidate cutoff frequency (taken in a logarithmically spaced grid of 200 values in the range $f_c \in [0.5,\,20]\,\mu\mathrm{m}^{-1}$, \citealp{Bidot_2024}).

Table \ref{tab:comparison_frequency} shows the cutoff frequencies obtained using this method ($f_c^{\star}$), along with the associated S/Ns. These are the values presented in Sect. \ref{subsec:Cross-correlation processing}, and we also give them in pixel$^{-1}$ (spectral pixel). Additionally, we show the cutoff frequency corresponding to the maximum S/N detected using the fake planet method (i.e., the fake planet with the highest S/N), $f_c^{\rm max}$, and the associated S/Ns. Finally, the table includes, for reference, the cutoff frequencies that maximize the S/N directly at the planet location (the first method described above, prior to the “fake planets” method), $f_c^{\rm opt}$, along with the S/N.

In addition to this study, we can also examine the impact of the cutoff frequency selection method on the S/N of the molecules using the molecular mapping method. We calculated the same maps as those presented in Sect. \ref{subsec:Molecular mapping}, but this time using the frequencies $f_c^{\rm max}$ and $f_c^{\rm opt}$. Table \ref{tab:comparison_molecule} presents these results. We observe that for the highest S/Ns (> 10), the S/Ns are better when using the optimization method on the actual planet (for \ch{NH3} in MIRI 3A S/N$^{\star}=14.4$ and S/N$^{\rm opt}=15.6$). This result is predictable, since these strongly detected molecules ($\ch{NH3}$ and $\ch{H2O}$) are the ones that yield high S/Ns in the detection maps. Thus, the optimization frequency is better suited for detecting these molecules than the average frequency of fake planets $f_c^{\star}$. However, we observe that for lower S/Ns, the S/N values are similar to or even better with $f_c^{\star}$ than with $f_c^{\rm opt}$, demonstrating the robustness of the method.

\begin{table*}
\centering
\caption{Comparison of S/N and cutoff frequency between the three methods: average frequency ($f_c^{\star}$, $\rm S/N^{\star}$), maximum frequency ($f_c^{\rm max}$, $\rm \rm S/N^{\rm max}$), and optimum frequency at the real planet's position ($f_c^{\rm opt}$, $\rm S/N^{\rm opt}$).}
\begin{tabular}{lcccccccc}
\hline\hline
Instrument & Range [$\mu$m] & $f_c^{\star}$ [$\mu\mathrm{m}^{-1}$] & $f_c^{\star}$ [10$^{-2}\,$pixel$^{-1}$] & S/N$^{\star}$ &  $f_c^{\rm max}$ [$\mu\mathrm{m}^{-1}$] & S/N$^{\rm max}$ & $f_c^{\rm opt}$ [$\mu\mathrm{m}^{-1}$] & S/N$^{\rm opt}$ \\
\hline
NIRSpec -  G395H  &2.87--5.27    &15.24 & 1.01  &13.9  &14.87 &14.5 &13.05 &15.2\\
MIRI - 1A         &4.90--5.74    &15.19 & 1.22  &10.2  &14.87 &10.2 &14.90 &10.6\\
MIRI - 1B         &5.66--6.63    &8.27  & 0.66  &0.4   &9.18  &0.4  &1.49  &0.9\\
MIRI - 1C         &6.53--7.65    &7.00  & 0.56  &7.5   &7.77  &7.8  &8.91  &8.0\\
MIRI - 2A         &7.51--8.77    &3.80  & 0.49  &3.4   &5.99  &2.3  &1.25  &4.1\\
MIRI - 2B         &8.67--10.13   &2.28  & 0.30  &5.6   &5.88  &2.2  &1.34  &6.0\\
MIRI - 2C         &10.01--11.70  &3.42  & 0.44  &6.4   &4.06  &5.9  &1.52  &6.6\\
MIRI - 3A         &11.55--13.47  &3.61  & 0.90  &13.8  &3.19  &14.3 &6.28  &14.8\\
MIRI - 3B         &13.34--15.57  &12.48 & 3.12  &9.0   &2.16  &7.7  &11.64 &9.2\\
MIRI - 3C         &15.41--17.98  &12.22 & 3.06  &15.0  &11.47 &15.1 &11.41 &15.1\\
MIRI - 4A         &17.70--20.95  &7.49  & 4.49  &3.5   &3.84  &3.4  &1.01  &4.3\\
MIRI - 4B         &20.69--24.48  &8.66  & 5.20  &0.3   &17.24 &0.0  &0.82  &3.6\\
MIRI - 4C         &24.40--27.90  &10.42 & 6.25  &0.0   &18.92 &0.7  &2.54  &1.6\\
\hline
\end{tabular}

\label{tab:comparison_frequency}
\end{table*}

\begin{table}[ht!]
\centering
\caption{Comparison of S/N from molecular mapping between the three methods: average frequency ($\rm S/N^{\star}$), maximum frequency ($\rm S/N^{\rm max}$), and optimum frequency at the real planet's position ($\rm S/N^{\rm opt}$).}
\begin{tabular}{lccccccc}
\hline\hline
Molecule & Instrument & S/N$^{\star}$ & S/N$^{\rm max}$ & S/N$^{\rm opt}$ \\
\hline
\ch{NH3}  &MIRI - 2C           &6.6   &7.4   &6.2\\
\ch{NH3}  &MIRI - 3A           &14.4  &14.0  &15.6\\
\ch{NH3}  &MIRI - 3B           &5.7   &7.9   &5.7\\
\ch{NH3}  &MIRI - 3C           &6.9   &6.4   &6.4\\
\ch{H2O}  &NIRSpec -  G395H    &10.8  &11.2  &10.8\\
\ch{H2O}  &MIRI - 1A           &9.7   &9.8   &8.7\\
\ch{H2O}  &MIRI - 3C           &11.3  &11.0  &11.0\\
\ch{HCN}  &MIRI - 3A           &5.3   &4.7   &5.6\\
\ch{PH3}  &MIRI - 2B           &4.1   &1.7   &4.0\\
\ch{CH4}  &NIRSpec -  G395H    &3.8   &3.7   &3.8\\
\ch{C2H2} &MIRI - 3C           &3.9   &3.8   &3.8\\
\ch{CO2}  &MIRI - 3C           &2.6   &2.5   &2.5\\
\hline
\end{tabular}

\label{tab:comparison_molecule}
\end{table}

\newpage

\section{S/N and spectrum}
\label{sec:S/N and spectrum}

\begin{figure}[t!]
    \centering
    \includegraphics[width=\linewidth]{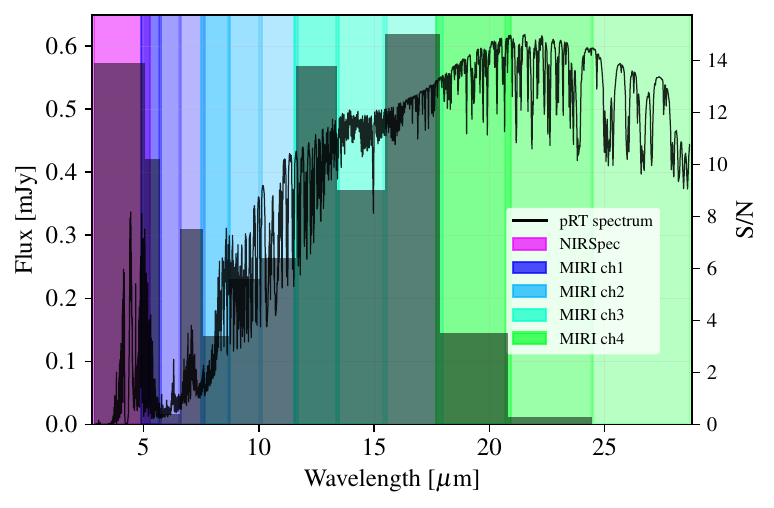}
    \caption{Comparison between the S/N as a function of wavelength (dark histogram) and the theoretical spectrum of the planet simulated using \texttt{petitRADTRANS}.
    }
    \label{fig:spectrum_vs_snr}
\end{figure}

It is important to note that the S/N measured at the planet's position through the cross-correlation method is directly related to the planet's spectrum. Indeed, Fig.~\ref{fig:spectrum_vs_snr} shows the S/N histogram overlaid with the petitRADTRANS model spectrum used for the detections. 
We note that the S/N drops to 0.4 in sub-channel 1B, compared to 10.2 in 1A and 7.5 in 1C. This is because the spectral range of sub-channel 1B corresponds to strong absorption, which results in a flux that is too weak to be detected (high opacity). CO2 could be responsible for this absorption, by being significantly stronger that shown in this model, making the band partly saturated, and leaving less signal to be probed. This would need to be confirmed with a future retrieval analysis.
Finally, an important observation is the overall drop in S/N in channel 4, with values of 2.5 in 4A, 0.3 in 4B, and 0.0 in 4C, notably lower than those obtained in channels 1 and 3. This is due to the loss of instrument sensitivity in channel 4, mainly caused by the telescope's thermal emission being stronger at these wavelengths \citep{Glasse_2015}. 
In addition, the instrument's spectral resolving power decreases from R $\gtrsim$ 3500 in channel 1 to R $\gtrsim$ 1500 in channel 4 \citep{Argyriou_2023}. Since the cross-correlation method relies on the ability to resolve individual spectral lines of the model, a lower spectral resolution spreads the line signal over more pixels and thus reduces the correlation efficiency, on top of the pure photometric sensitivity loss. These two effects therefore combine to explain the marked drop in S/N observed in channel 4.

\end{appendix}

\end{document}